# Sequential Data Mining using Correlation Matrix Memory

[a]Sanil Shanker KP, [b]Aaron Turner, [c]Elizabeth Sherly, [b]Jim Austin

[a]Department of Computer Science  [b]Department of Computer Science  [c]Indian Institute of Information Technology
University of Kerala , India         University of York, UK              and Management- Kerala, India
e-mail: sanilshankerkp@gmail.com

**Abstract:** This paper proposes a method for sequential data mining using correlation matrix memory. Here, we use the concept of the Logical Match to mine the indices of the sequential pattern. We demonstrate the uniqueness of the method with both the artificial and the real datum taken from NCBI databank.

## I  INTRODUCTION

In this paper we put forward a unique method of search for sequential patterns using correlation matrix memory by Logical Match. The correlation matrix memory is a particular type of binary associative neural network (Austin,1994). In Logical Matching strategy, the sequence is arranged so that each element in the pattern coincides with it's corresponding index and then proceed to match logically the indices of the subsequence with those of the text pattern (Sanil et al,2010). A correlation matrix memory can be represented as a m X n binary matrix D, where m and n are input width and output width respectively. For a given sequential pattern the input binary vector $I_{SP}$ operates logical AND with reference matrix (database D) gives $I_{SP}^T$. During the transformation, $O_P \leftarrow I_P \cdot D$, we mine the indices of both the pattern and text using logical matching. where $I_{SP}^T = O_{SP,}$ the output binary vector of the sequential pattern .

## II  METHOD

1.Initialize Text and Pattern

2.Initialize input binary vector $I_P \leftarrow$ Pattern and $I_T \leftarrow$ Text

3.Create the reference database D

4. Select (indices of Pattern)  when  $O_P \leftarrow I_P \cdot D$

5. Select (indices of Text)  when  $O_T \leftarrow I_T \cdot D$

6. Match (indices of Pattern) with (indices of Text) in the increasing order of indices.

## III  SIMULATION WITH ARTIFICIAL DATUM

Text => <CTCACTCCTC>

Pattern => <CTC>

Initialize {0001← A, 0010← T, 0100← G, 1000← C}

Shift the Text(Table 1) so that the input binary vector $I_{Text}$ operates logical AND with reference matrix ( Database D) gives $I_{Text}^T$ (Table 2).

Input Text ($I_{Text}$)                                      Database D

C  T  C  A  C  T  C  C  T  C  **AND**  0/1  0/1  0/1  0/1

| 1 | 0 | 1 | 0 | 1 | 0 | 1 | 1 | 0 | 1 | . | 0 | 0 | 0 | 1 |
|---|---|---|---|---|---|---|---|---|---|---|---|---|---|---|
| 0 | 0 | 0 | 0 | 0 | 0 | 0 | 0 | 0 | 0 | . | 0 | 0 | 1 | 0 |
| 0 | 1 | 0 | 0 | 0 | 1 | 0 | 0 | 1 | 0 | . | 0 | 1 | 0 | 0 |
| 0 | 0 | 0 | 1 | 0 | 0 | 0 | 0 | 0 | 0 | . | 1 | 0 | 0 | 0 |

Table 1.                                      ↓   ↓   ↓   ↓

Table 2.  Transpose of Input Text $I_{Text}^T$       w   x   y   z

| Indices | w | x | y | z |
|---|---|---|---|---|
| 10 | 0 | 0 | 0 | 1 |
| 9 | 0 | 1 | 0 | 0 |
| 8 | 0 | 0 | 0 | 1 |
| 7 | 1 | 0 | 0 | 0 |
| 6 | 0 | 0 | 0 | 1 |
| 5 | 0 | 1 | 0 | 0 |
| 4 | 0 | 0 | 0 | 1 |
| 3 | 0 | 0 | 0 | 1 |
| 2 | 0 | 1 | 0 | 0 |
| 1 | 0 | 0 | 0 | 1 |

That is, indices of the Text, <w(7);

x(2,5,9);

z(1,3,4,6,8,10)>

Shift the Pattern (Table 3) so that the input binary vector $I_{Pattern}$ operates logical AND with reference matrix (Database D) gives $I_{Pattern}^T$ (Table 4).

Input Pattern ($I_{Pattern}$)    Database D

| C | T | C | **AND** | 0/1 | 0/1 | 0/1 | 0/1 |
|---|---|---|---|---|---|---|---|
| 1 | 0 | 1 | . | 0 | 0 | 0 | 1 |
| 0 | 0 | 0 | . | 0 | 0 | 1 | 0 |
| 0 | 1 | 0 | . | 0 | 1 | 0 | 0 |
| 0 | 0 | 0 | . | 1 | 0 | 0 | 0 |

Table 3.   ↓   ↓   ↓   ↓

w   x   y   z

Table 4. Transpose of Input Pattern $I_{Pattern}^T$

| Indices | w | x | y | z |
|---|---|---|---|---|
| 3 | 0 | 0 | 0 | 1 |
| 2 | 0 | 1 | 0 | 0 |
| 1 | 0 | 0 | 0 | 1 |

That is, indices of the pattern, <x(2);

z(1,3)>

Locating Pattern in the Text:

z: <u>1</u>  3  <u>4</u>  6  <u>8</u>  10

x: <u>2</u>  <u>5</u>  <u>9</u>

z: 1  <u>3</u>  4  <u>6</u>  8  <u>10</u>

Here, Pattern CTC is repeating in the locations (1,2,3);(4,5,6);(8,9,10) of the text.

### IV  EXPERIMENTAL RESULTS

For testing the proposed method, the program has been written in C++ language under Linux platform. The method was tested against DNA sequences of various sizes taken from NCBI databank(Table 5). The method provides the solution to problem of locating the exact position of the pattern in the text.

Table 5.    Location of Pattern in Text

| | |
|---|---|
| Kennedy disease<br><br>Gene: AR<br><br>Repeat motif: CAG<br><br>Locus: NM_000044<br><br>Range: 1-700 | (16,17,18);(53,54,55);<br>(59,60,61);(71,72,73);<br>(101,102,103);(111,112,113);<br>(136,137,138);(142,143,144);<br>(153,154,155);(159,160,161);<br>(175,176,177);(363,364,365);<br>(372,373,374);(386,387,388);<br>(441,442,443);(450,451,452);<br>(463,464,465);(533,534,535);<br>(554,555,556);(562,563,564);<br>(663,664,665);(685,686,687);<br>(689,690,691);(697,698,699) |
| Huntington disease<br><br>Gene: HD<br><br>Repeat motif: CAG<br><br>Locus: NM_002111<br><br>Range: 1-510 | (33,34,35);(57,58,59);<br>(196,197,198);(199,200,201);<br>(202,203,204);(205,206,207);<br>(208,209,210);(211,212,213);<br>(214,215,216);(217,218,219);<br>(220,221,222);(223,224,225);<br>(226,227,228);(229,230,231);<br>(232,233,234);(235,236,237);<br>(238,239,240);(241,242,243);<br>(244,245,246);(247,248,249);<br>(250,251,252);(253,254,255);<br>(256,257,258);(262,263,264);<br>(298,299,300);(307,308,309);<br>(319,320,321);(325,326,327);<br>(340,341,342);(346,347,348);<br>(428,429,430);(487,488,489);<br>(495,496,497);(506,507,508) |
| Friedreich ataxia<br><br>Gene: FRDA<br><br>Repeat motif: GAA<br><br>Locus: AH003505S1<br><br>Range: 1- 2465 | (86,87,88);(151,152,153);<br>(158,159,160);(232,233,234);<br>(383,384,385);(391,392,393);<br>(491,492,493);(508,509,510);<br>(549,550,551);(816,817,818);<br>(947,948,949);(1011,1012,1013);<br>(1015,1016,1017);(1065,1066,1067);<br>(1300,1301,1302);(1304,1305,1306);<br>(1318,1319,1320);(1327,1328,1329);<br>(1345,1346,1347);(1378,1379,1380);<br>(1485,1486,1487);(1535,1536,1537);<br>(1558,1559,1560);(1601,1602,1603);<br>(1631,1632,1633);(1639,1640,1641);<br>(1649,1650,1651);(1668,1669,1670);<br>(1711,1712,1713);(1723,1724,1725);<br>(1774,1775,1776);(1781,1782,1783);<br>(1875,1876,1877);(1976,1977,1978);<br>(1979,1980,1981);(2022,2023,2024);<br>(2088,2089,2090);(2140,2141,2142);<br>(2184,2185,2186);(2187,2188,2189);<br>(2190,2191,2192);(2193,2194,2195);<br>(2196,2197,2198);(2199,2200,2201);<br>(2202,2203,2204);(2205,2206,2207);<br>(2208,2209,2210);(2217,2218,2219);<br>(2282,2283,2284);(2427,2428,2429) |

## V    CONCLUSION

We present a new sequential data mining method using correlation matrix memory. Here, we use the concept of Logical Match to locate the pattern in the text. This method can possibly be implement to develop a new approach related to the sequential data mining.

### ACKNOWLEDGEMENT

SSKP was funded in part by European Research and Educational Collaboration with Asia